\PassOptionsToPackage{dvipsnames}{xcolor}

\documentclass[sigplan,screen,nonacm]{acmart}

\usepackage[english]{babel}
\usepackage{multirow}
\usepackage{csquotes}
\usepackage{adjustbox}
\usepackage{wrapfig}
\usepackage[compatibility=false]{caption}
\usepackage{subcaption}
\usepackage{pifont}
\usepackage{amsmath}
\usepackage{mathbbol}
\usepackage[utf8]{inputenc}
\usepackage{lipsum}
\usepackage{xparse}
\usepackage{mathtools}
\usepackage{amsthm}
\usepackage{collcell}
\usepackage{mathpartir}
\usepackage{hyperref}
\usepackage{cleveref}
\usepackage{wrapfig}

\usepackage{listings}
\NewDocumentCommand{\fresh}{}{\operatorname{fresh}}
\NewDocumentCommand{\CGEnv}{}{\Gamma}

\theoremstyle{definition}

\NewDocumentCommand{\efmt}{m}{\textbf{\texttt{#1}}}
\NewDocumentCommand{\class}{}{\efmt{class}}
\NewDocumentCommand{\interface}{}{\efmt{interface}}

\NewDocumentCommand{\superfg}{}{\texttt{super}}

\NewDocumentCommand{\ethis}{}{\texttt{this}}

\NewDocumentCommand{\extends}{}{\ensuremath{\mathrel{\efmt{extends}}}}
\NewDocumentCommand{\implements}{}{\ensuremath{\mathrel{\efmt{implements}}}}

\NewDocumentCommand{\idfmt}{m}{\textbf{#1}}
\NewDocumentCommand{\aeither}{}{\textit{A}}

\NewDocumentCommand{\aid}{}{\idfmt{A}}

\NewDocumentCommand{\ahole}{}{\idfmt{?A}}

\NewDocumentCommand{\feither}{}{\textit{M}}

\NewDocumentCommand{\meither}{}{\textit{M}}
\NewDocumentCommand{\methid}{}{\idfmt{M}}
\NewDocumentCommand{\mhole}{}{\idfmt{?M}}

\NewDocumentCommand{\tvar}{}{\alpha}

\NewDocumentCommand{\tmethod}{}{t}

\NewDocumentCommand{\mif}{}{\textbf{if }}
\NewDocumentCommand{\mwhere}{}{\textbf{where }}

\DeclarePairedDelimiterX\program[1]\lbrace\rbrace{#1}

\newcommand{\super}{\operatorname{\textsf{super}}}

\newcommand{\interfaces}{\operatorname{\textsf{interfaces}}}
\newcommand{\classes}{\operatorname{\textsf{classes}}}

\NewDocumentCommand{\ol}{}{\overline}
\NewDocumentCommand{\Lm}{}{\ensuremath{\mathcal{L}_{M}}}
\NewDocumentCommand{\Prog}{}{\ensuremath{P}}
\NewDocumentCommand{\Constraints}{}{\ensuremath{C}}

\NewDocumentCommand{\texists}{}{\textsf{exists}}
\NewDocumentCommand{\timpl}{}{\textsf{implements}}

\NewDocumentCommand{\matcha}{}{\dashleftarrow}
\NewDocumentCommand{\matchm}{}{\dashleftarrow}
\NewDocumentCommand{\patc}{}{\ensuremath{\mathcal{A}}}
\NewDocumentCommand{\patm}{}{\ensuremath{\mathcal{M}}}

\NewDocumentCommand{\pcat}{}{\oplus}

\DeclarePairedDelimiterX{\Struct}[1]{\lbrack}{\rbrack}{\IfNoValueTF{#1}{\cdot}{#1}}

\NewDocumentCommand{\fsec}{m}{\rotatebox[]{90}{\textbf{\footnotesize #1}}}

\newcolumntype{V}{>{\collectcell\fsec}r<{\endcollectcell}}
\newcolumntype{B}{>{$}l<{$ \vspace{1em}}}

\NewDocumentCommand{\tsmall}{m}{\ensuremath{\footnotesize #1}}
\newcolumntype{s}{>{\collectcell\tsmall}r<{\endcollectcell}}

\DeclarePairedDelimiter{\Suspend}{\Lparen}{\Rparen}
\NewDocumentCommand{\Constraint}{o m}{\ensuremath{\IfValueTF{#1}{\overset{\textsc{#1}}{#2}}{#2}}}
\NewDocumentCommand{\Suspended}{o m}{\ensuremath{\IfValueTF{#1}{\overset{\textsc{#1}}{\Suspend{#2}}}{\Suspend{#2}}}}

\NewDocumentCommand{\Fragments}{}{\ensuremath{F}}

\NewDocumentEnvironment{ constraints }{O{rcl}}
  {\begin{array}{#1}}
  {\end{array}}

\NewDocumentCommand \becomesSymbol{s}{\IfBooleanTF{#1}{\twoheadrightarrow^*}{\twoheadrightarrow}}%
\NewDocumentCommand \synthSymbol{}{\rightsquigarrow}

\NewDocumentCommand \PS{o}{\IfValueTF{#1}{\left\langle \Prog,~ \Constraints \land #1 \right\rangle}{\left\langle \Prog,~ \Constraints \right\rangle}}
\NewDocumentCommand \PSf{O{_f}}{\left\langle \Prog#1,~ \Constraints#1 \right\rangle}
\NewDocumentCommand \con {s m}{\Prog \vdash \Constraints \land #2 & \becomesSymbol}
\NewDocumentCommand \fails{}{& \bot}%
\NewDocumentCommand \becomes{o}{& \IfValueTF{#1}{\Constraints \land #1}{\Constraints}}
\NewDocumentCommand \cexp{s m}{\IfBooleanT{#1}{\\ &}& #2}
\NewDocumentCommand \cif{s m}{& \IfBooleanT{#1}{\\ &}\ \mif #2}
\NewDocumentCommand \cwhere{s m}{& \IfBooleanT{#1}{\\ & &}\ \mwhere #2}
\NewDocumentCommand \cotherwise{s O{otherwise}}{& \IfBooleanT{#1}{\\ & &}\ \text{#2}}
\NewDocumentCommand \subtyp{}{\leq}

\NewDocumentCommand{\FigSym}{m}{\hfill\framebox{\small#1}}

\NewDocumentCommand{\Mason}{}{\textsc{Mason}}
\NewDocumentCommand{\Foo}{}{\texttt{Foo}}

\hypersetup{colorlinks=true, allcolors=BrickRed}

\acmJournal{PACMPL}

\title{Mason: Type- and Name-Guided Program Synthesis}

\author{Jasper Geer}
\orcid{0009-0006-6839-2305}
\affiliation{%
  \institution{University of British Columbia}
  \city{Vancouver}
  \country{Canada}}
\email{jasper.geer@ubc.ca}

\author{Fox Huston}
\orcid{0000-0002-9290-0203}
\affiliation{%
  \institution{Tufts University}
  \city{Medford}
  \country{USA}}
\email{hello@fox.boston}

\author{Jeffrey S. Foster}
\orcid{0000-0001-8043-1166}
\affiliation{%
  \institution{Tufts University}
  \city{Medford}
  \country{USA}}
\email{jeffrey.foster@tufts.edu}

\begin{document}

\begin{abstract}

  Object-oriented programs tend to be written using many common coding idioms, such as those captured by design patterns.  While design patterns are useful, implementing them is often tedious and repetitive, requiring boilerplate code that distracts the programmer from more essential details.
  In this paper, we introduce \Mason{}, a tool that synthesizes object-oriented programs from partial program pieces, and we apply it to automatically insert design patterns into programs.
  At the core of \Mason{} is a novel technique we call \emph{type- and name-guided synthesis}, in which an enumerative solver traverses a partial program to generate typing constraints; discharges constraints via program transformations guided by the names of constrained types and members; and backtracks when a constraint is violated or a candidate program fails unit tests.
  We also introduce two extensions to \Mason{}: a non-local backtracking heuristic that uses execution traces, and a language of patterns that impose syntactic restrictions on missing names.
  We evaluate \Mason{} on a suite of benchmarks to which \Mason{} must add various well-known design patterns implemented as a library of program pieces.
  We find that \Mason{} performs well when very few candidate programs satisfy its typing constraints and that our extensions can improve \Mason{}'s performance significantly when this is not the case.
  We believe that \Mason{} takes an important step forward in synthesizing multi-class object-oriented programs using design patterns.

\end{abstract}

\settopmatter{printfolios=true} 
\bibliographystyle{ACM-Reference-Format}
\maketitle
\sloppy


\section{Introduction}

\begin{figure*} [t]
  \begin{center}
  \begin{tabular}{ p{0.5\linewidth} p{0.5\linewidth} }
  \begin{lstlisting}
class Canvas extends Object {
  Integer brushX;
  Integer brushY;
  Image image;
  ... // other fields
  Canvas(...) { ... }
  Image currImage() { ... }
  void startDrawing() { ... }
  void stopDrawing() { ... }
  void moveBrush(Integer x, Integer y) {
    ...
    this.setBrushX(this.brushX + x);
    this.setBrushY(this.brushY + y);
    ...
}  }
  \end{lstlisting}
    &
  \begin{lstlisting}
class CanvasLogger {
  List log = new ArrayList();
  // state-change handlers
  void onUpdateX(Integer newX) {
    this.log.add("x: "
      .concat(newX.toString()));
  }
  void onUpdateY(Integer newY) {
    this.log.add("y: "
      .concat(newY.toString()));
  }
  // other methods
  getLogEntry(Integer i) {
    return this.log.get(i);
}  }
  \end{lstlisting}
  \end{tabular}
  \end{center}
  \caption{Partial Program Input to \Mason{}}
  \label{fig:canvas-def}
\end{figure*}

Users of object-oriented programming languages often use design patterns, recipes for common idioms in software engineering~\cite{gamma1994}.
Design patterns aid in building more maintainable software and facilitate communication between software engineers.
However, many design patterns require the programmer to lay down large amounts of uninteresting boilerplate code before they can focus on essential details.

A potential solution to this problem is program synthesis, which seeks to make programmers more productive by automatically generating code from user-provided specifications.
Prior work in synthesis for object-oriented languages is capable of leveraging lightweight formal specifications to synthesize methods in Ruby~\cite{guria2021, guria2023}, synthesizing programs via sketching~\cite{solarlezama09} in Java~\cite{jeon2015, mariano19}, and generating mock library implementations~\cite{jeon2016} for program analysis.
However, such techniques are either highly domain specific or synthesize low-level implementation details while requiring the programmer to supply the boilerplate.

In this paper, we introduce \Mason{}, a new tool for synthesizing object-oriented programs, and we show how to use it to automatically insert design patterns into programs.
The key novelty of \Mason{} is that it synthesizes object-oriented programs from partial program pieces guided by both types and names.
In \Mason{}, the programmer provides a partial candidate program, a set of unit tests it must pass, and a library of partial program pieces that describe a design pattern.
We call these program pieces \emph{fragments}, and they may include \emph{holes} in place of type or member names.
\Mason{} begins search by traversing the candidate program to generate typing constraints.
Then, \Mason{} solves typing constraints in one of two ways: unifying the name of a constrained type or member with a name in the candidate program by filling a hole, or synthesizing a definition for a missing type or member by merging a fragment into the candidate program.
After each transformation, \Mason{} rechecks constraints and backtracks if any are contradicted.
Once all goals are discharged, the candidate program is run and \Mason{} backtracks if any unit tests fail.

We formalize \Mason{} for a core object-oriented language \Lm{} which permits holes in place of type and member names.
We define a type inference algorithm for \Lm{} in two parts:
first, a syntax-directed procedure for generating typing constraints from \Lm{} programs;
second, a constraint rewriting system that checks and simplifies constraints in the context of a candidate program.
Programs may be partial, so some constraints may be unsatisfied because they constrain undefined types or members; we call these \emph{suspended}.
We then define the rules of \emph{type- and name-guided synthesis} which solves suspended constraints by composing fragments and filling holes with concrete names.

We implement \Mason{} in Scala as a tool that accepts partial Java programs with holes.
We adapt our non-deterministic synthesis rules to a search procedure by imposing an ordering on suspended constraints.
We also implement two extensions that prune candidates that cannot be eliminated using typing constraints:
\emph{Trace-guided backtracking} is a non-local backtracking heuristic that uses the execution traces of failing unit tests and \emph{name patterns}, which may be used to decorate holes with constraints on the set of names that may fill them.

We evaluate \Mason{} and its extensions using a suite of benchmarks that we developed in which \Mason{} must apply Gang of Four~\cite{gamma1994} design patterns implemented as a library of fragments.
Each benchmark is scalable relative to a natural number, and we measure \Mason{}'s time to find a correct solution at increasing scales.
We find that baseline \Mason{} performs well in benchmarks where the number of well-typed synthesizable programs remains low.
For all but one of the remaining benchmarks, trace-guided backtracking allows \Mason{} to synthesize larger programs before the timeout, and name patterns reduce times from above 120 seconds to below 6 seconds for two benchmarks.

We believe that \Mason{} takes an important step forward in synthesizing multi-class object-oriented programs using design patterns.

\section{Motivating Example}%
\label{sec:overview}

We now use \Mason{} to complete a partial Java
program by implementing the Observer~\cite{gamma1994} design pattern.
\Mason{} is provided with an \emph{initial} partial program piece along with a library
of program pieces, and composes a solution guided by types, names, and unit tests.

Consider the definitions of \texttt{Canvas} and \texttt{CanvasLogger} shown in \Cref{fig:canvas-def}. \texttt{Canvas} is a class that allows a client to manipulate an image with a virtual brush.
Instances of \texttt{CanvasLogger} record the movements of said brush using two handlers: \texttt{onUpdateX} and \texttt{onUpdateY}.
We also require that a single \texttt{Canvas} be able to notify multiple subscribed \texttt{CanvasLogger}s.
This is a common kind of dependency which is addressed by the Observer~\cite{gamma1994} design pattern.

In the Observer pattern, \emph{subject} objects carry a set of \emph{observers}.
The subject is responsible for notifying each of its observers whenever a meaningful change in state occurs.
In the version of the pattern we intend to synthesize, certain state-changing methods of the subject will be responsible for notifying any subscribed observers of the change.

In concrete terms, we envision a special setter method for each of
\texttt{brushX} and \texttt{brushY}, as well as a method for registering new
\texttt{CanvasLogger}s to a \texttt{Canvas}. To communicate this intent to
\Mason{}, we provide a \emph{fragment library}, a set of program pieces from
which \Mason{} may derive new type and method declarations, as well as a set of
unit tests which the synthesized program must pass.

\paragraph{Fragments} One of these program pieces, or \emph{fragments}, is shown in \Cref{fig:example-frag}.
The \texttt{@MemberFragment} annotations indicates that we intend \Mason{} to synthesize member implementations using this fragment.
The identifiers prefixed with \texttt{?} on lines \ref{line:classname}, \ref{line:methname}, \ref{line:for} \ref{line:field}, \ref{line:obs}, and \ref{line:notify} are \emph{holes}, syntactic metavariables that \Mason{} is expected to fill.
Holes indicate which details \Mason{} is allowed to manipulate to adapt a fragment into a new context.
For example, the class name, \texttt{?Update}, will be flled with the name of an existing class, and their declarations merged.
From the structure of the method definition we also see that this fragment will be used to synthesize one of our setter methods, updating some field \texttt{?field} with a new value of some type \texttt{?T}, and calling some handler \texttt{?notify}, on each \texttt{?Observer} to notify it of the change.
By passing \texttt{newValue} to \texttt{?notify}, we communicate a key detail about this design pattern's shape: each state-changing method of our subject is mapped to an particular handler of the observer.

\begin{figure}[ht]
\begin{lstlisting}
@MemberFragment
class ?@?Update@? { *@\label{line:classname}@*
   void ?@?{(set)(?field)}@?(?@?T@? newValue) { *@\label{line:methname}@*
     this.?@?field@? = newValue; *@\label{line:field}@*
     for (int i = 0; i < this.?@?{(num)(?)}@?(); i++) { *@\label{line:for}@*
       ?@?Observer@? o = this.?@?{(get)(?)}@?(i)); *@\label{line:obs}@*
       o.?@?notify@?(newValue); *@\label{line:notify}@*
}  }  }
\end{lstlisting}
\caption{Example Element of Fragment Library Provided to \Mason{}. Holes begin with ? and are written in \textcolor{red}{\textbf{red}}.}
\label{fig:example-frag}
\end{figure}

\paragraph{Constraint Generation} \Mason{} traverses the initial fragment, collecting \emph{typing constraints}.
Notice that the implementation of \texttt{Canvas} in \Cref{fig:canvas-def} is incomplete.
Specifically, the body of \texttt{moveBrush} references undefined methods \texttt{setBrushX} and
\texttt{setBrushY}.
Instead of signalling failure, however, \Mason{} emits two \emph{suspended} typing constraints:

\begin{center}
  $\Suspend{\texttt{Canvas} \subtyp [ \texttt{setBrushX} : (\texttt{Integer}) \rightarrow \alpha ]}$\\
  $\Suspend{\texttt{Canvas} \subtyp [ \texttt{setBrushY} : (\texttt{Integer}) \rightarrow \beta ]}$
\end{center}

Informally, each constraint is satisfied if \texttt{Canvas} defines a method
with the given name from \texttt{Integer} to some type, represented by the type variables $\alpha$ and $\beta$.
The delimiters $\Suspend{\_}$ denote that while neither is satisfied, they each contain some undefined part that we may synthesize at some later step.
In this case, neither \texttt{setBrushX} nor \texttt{setBrushY} is defined.
We observe that constraints like these correspond to synthesis goals; we must synthesize implementations for both \texttt{setBrushX} and \texttt{setBrushY} before we can produce an executable program.
\Mason{} searches the fragment library for a method of identical arity and compatible type and may choose, for example, the fragment from \Cref{fig:example-frag}.

Note the unique syntax of the method name \texttt{?\{(set)(?field)\}}.
To provide a more detailed specification to \Mason{}, we have imposed a naming convention with the use of a \emph{name pattern}.
This pattern enforces that any name used to fill the hole, e.g., \texttt{setBrushX}, must consist of \texttt{set} followed by the name found in place of \texttt{?field}, modulo case.
However, \texttt{?field} is not yet known. To ensure that whichever name fills \texttt{?field} is consistent with the pattern, \Mason{} emits a \emph{name constraint}:

\begin{center}
  $\texttt{setBrushX} \matcha \texttt{(set)(?field)}$
\end{center}

The left operand of the constraint is a name and the right operand is a name pattern.
The constraint is satisfied if its left operand is concrete and matches the pattern on the right; holes like \texttt{?field} are treated as wildcards.
We describe name patterns in detail in \Cref{sec:extensions}.
\Mason{} does not use name constraints as synthesis goals but they can be highly effective at pruning the search space, as we will describe shortly.

\begin{figure*} [ht]
  \includegraphics*[width=14cm]{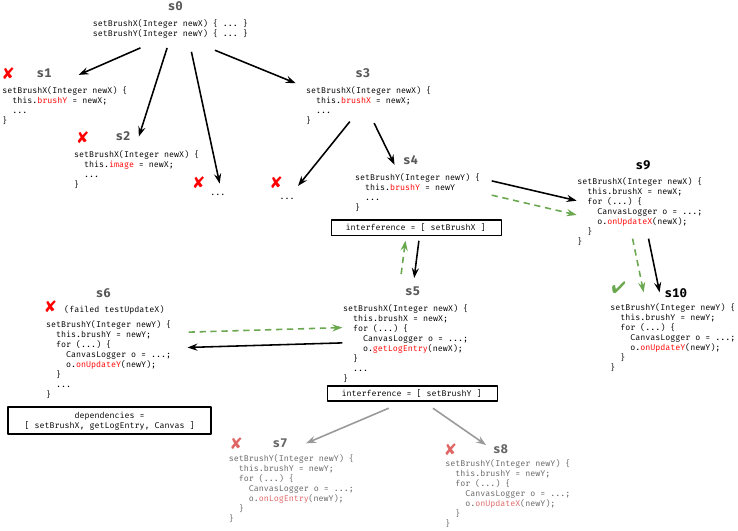}
  \caption{
    \Mason{}'s search for the correct Observer pattern implementation.
    Nodes denote synthesis steps and edges denote a particular choice taken at a given synthesis step.
    Each node is annotated with the candidate program before the synthesis step is taken and the interference or dependency sets of select steps are given in a box below.
    Leaves annotated with \textcolor{red}{\ding{56}} are candidates rejected for failing constraint resolution, or for failing unit tests.
    \textcolor{lightgray}{\textbf{gray}} nodes and edges are paths left unexplored by trace-guided backtracking.
    The \textcolor{ForestGreen}{\textbf{green}} dotted arrows trace the path of the search under trace-guidance.
  }
  \label{fig:observer-search}
\end{figure*}

\paragraph{Synthesis} \Mason{} searches for a program that satisfies all type and name constraints via a depth-first backtracking search, which is visualized in \Cref{fig:observer-search}.
Each element of the search space is characterized by a candidate program and a set of constraints.
\Mason{} first performs \emph{constraint resolution}, rewriting constraints in the context of the candidate program.
Constraint resolution has two purposes:
first, to typecheck partial programs, failing if subsequent steps could not possibly result in a well-typed program;
second, to update \Mason{}'s synthesis goals by discharging any solved constraints and suspending any eligible constraints.
If constraint resolution fails, \Mason{} backtracks.
We see this in \texttt{s2}, where field \texttt{image} of type \texttt{Image} is assigned \texttt{newX} of type \texttt{Integer}; constraint resolution fails and \Mason{} backtracks to \texttt{s0}.
If constraint resolution succeeds, \Mason{} chooses a suspended constraint to solve and explores possible solutions by applying \emph{synthesis rules}, which produce new candidate programs by filling holes or synthesizing definitions.

Suppose that \Mason{} has instantiated the fragment from \Cref{fig:example-frag} as the body for both \texttt{setBrushX} and \texttt{setBrushY} and correctly mapped \texttt{?Observer} to \texttt{CanvasLogger}, \texttt{?\{(num)(?)\}} to \texttt{numLoggers}, and \texttt{?\{(get)(?)\}} to \texttt{getLogger}.
This state corresponds to \texttt{s0} in \Cref{fig:observer-search}.
\Mason{} first explores $s1$, filling \texttt{?field} with \texttt{brushY} in the body of \texttt{setBrushX} to solve a suspended constraint $\Suspend{\texttt{Canvas} \subtyp [ \texttt{?field} : \gamma ]}$.
However, name constraints are renamed with the rest of the program, and this substitution transforms the previously mentioned constraint into $\texttt{setBrushX} \matcha \texttt{(set)(brushY)}$, which fails constraint resolution, causing \Mason{} to reject \texttt{s1}.
Constraint resolution also fails for \texttt{s2} and a series of elided candidates.
Eventually, \Mason{} reaches the only valid choice, \texttt{brushX}, and we arrive at \texttt{s3}.
Here too, only a single choice is valid and we inevitably arrive at \texttt{s4}.

\paragraph{Trace-Guided Backtracking}
When static information does not sufficiently narrow the search space, \Mason{} relies on dynamic information provided by unit tests to perform non-local backtracking.
Below is one such test, which we have provided to \Mason{} to specify the behavior of the solution.
Referring to \Cref{fig:canvas-def}, this test will pass only if \texttt{setBrushX} calls \texttt{onUpdateX}.

\begin{center}
\begin{lstlisting}
@Test
static void testUpdateX() {
  CanvasLogger cl = new CanvasLogger();
  Canvas c = new Canvas(...);
  c.registerLogger(cl);
  assert(c.getLogger(0) == cl);
  assert(c.numLoggers() == 1)
  c.setBrushX(10);
  assert(cl.log.contains("x: 10"));
}
\end{lstlisting}
\end{center}

Three well typed options exist to fill \texttt{?notify}: \texttt{onUpdateX},
\texttt{onUpdateY}, and \texttt{getLogEntry}. Suppose that \Mason{} makes the
wrong choice, picking \texttt{getLogEntry} and arriving in \texttt{s5}.
Despite correctly completing the body of \texttt{setBrushY}, \texttt{s6} fails test
\texttt{testUpdateX}. Under a naive backtracking strategy, \Mason{} might
attempt different \texttt{setBrushY} implementations with \texttt{s7} and
\texttt{s8}, but this is will be fruitless; \texttt{setBrushY} is never invoked
during the execution of \texttt{testUpdateX}.

\Mason{} mechanizes this intuition as \emph{trace-guided backtracking}.
For each synthesis step, \Mason{} computes the \emph{interference set} of methods whose implementation the step may affect.
Then, for each test, \Mason{} tracks a \emph{dependency set} of methods upon which the outcome of the test might depend.
If a synthesis step's interference set intersects with the dependency set of a unit test, then \Mason{} has determined that the synthesis step may affect the outcome of that test.

Upon synthesizing a complete, well-typed candidate program, \Mason{} discovers dependencies by executing unit tests sequentially, short-circuiting on  the first failing test.
\Mason{} records the set of methods that were invoked during the execution of this last unit test, its \emph{execution trace}, and adds these to the test's dependency set.
Then, \Mason{} backtracks non-locally until reaching a synthesis step whose interference set intersects with the dependency set of the failing test, selects an unvisited candidate to explore, and replays any choices from the previous path that it can.

We observe this in \Cref{fig:observer-search}.
Because the choice made at the outgoing edges of \texttt{s4} affects, or \emph{interferes} with the implementation of \texttt{setBrushX}, it is added to \texttt{s4}'s interference set (see \Cref{sec:extensions} for a definition of interference).
Likewise, the interference set of \texttt{s5} includes \texttt{setBrushY}.
After executing the program at \texttt{s6}, \texttt{testUpdateX} fails
with \texttt{setBrushX}, \texttt{getLogEntry}, and \texttt{Canvas} as
dependencies, since these were the function calls made prior to the failure.

After backtracking to \texttt{s5}, \Mason{} backtracks again, skipping \texttt{s7} and \texttt{s8} because \texttt{s5} interferes only with \texttt{setBrushY}, which \texttt{testUpdateX} does not depend on.
After reaching \texttt{s4}, \Mason{} correctly chooses \texttt{onUpdateX} and replays $(\texttt{s5}, \texttt{s6})$, to find the solution, \texttt{s10}.
In this way, \Mason{} prunes large swathes of the search space by only reevaluating choices which may have contributed to a test failure.
In this scenario, there are $3^{2}$ possible executable candidates reachable from \texttt{s4}.
With trace-guided backtracking we visit at most $3 \times 2$.

\section{Formalism}
\label{sec:formalism}

In this section, we define an object-oriented core language \Lm{} to formalize
type- and name-guided program synthesis.
The top portion of \Cref{fig:syntax} shows the syntax of \Lm{}.
Here, the notation $\ol{a}$ indicates a sequence of zero or more occurrences of $a$.
Programs consist of type declarations $d$, each defining a class or interface.
Classes contain methods $m$ and fields~$f$, while interfaces contain
method signatures $s$.
Every class has a superclass and implements zero or more interfaces.

The bodies of \Lm{} methods consist of a single expression~$e$, which may be a variable reference $x$, field access $e.\meither$, method call $e.M(e)$, or type cast $(A)\ e$.
We omit forms such as assignment and sequencing but it is straightforward to extend our system to include such features.
Type and method names may either be \emph{concrete names} \aid{} and \methid{} for classes and methods, respectively, or \emph{holes} \ahole{} and \mhole{}, which are replaced by concrete names during synthesis.

\begin{figure}[t]
  \centering
	\begin{displaymath}
		\begin{array}{rcl}
    \Prog     & ::= & \ol{d} \\
    d     & ::=   & \class\ \aeither \extends \aeither \implements \ol{\aeither}\ \{\ \ol{f}\ \ol{m}\ \} \\
                      & \mid  & \interface\ \aeither\ \{\ \ol{s} \ \} \\
    m     & ::=   & \aeither\ \meither (\aeither\ x)\ \{\ e\ \} \\
    s     & ::=   & \aeither\ \meither (\aeither\ x) \\
    f     & ::=   & \aeither\ \feither \\
    e    & ::=    & x\
                                \mid\ e.\meither\
                                \mid\  e.M(e)
                                \mid\ (\aeither) e \\
    \aeither & ::= & \aid \mid \ahole \\
    \meither & ::= & \methid \mid \mhole \\

	\\

    t & ::= & \aeither
        \mid \tvar
        \mid t \rightarrow t \\
    \Constraints & ::=  & \Constraints \land \Constraints
        \mid t \subtyp t
        \mid t \subtyp \Struct{\meither:\tmethod}
        \mid \timpl(\aeither, \aeither) \\
    & \mid & \texists(\aeither)
        \mid \Suspend{ \Constraints }

		\end{array}
	\end{displaymath}
  \caption{\Lm{} Syntax, Types, and Constraints.}
  \label{fig:syntax}
\end{figure}
\paragraph{Typing Constraints}
Typechecking \Lm{} programs consists of generating and resolving typing constraints, the syntax of which is given in the bottom portion of \Cref{fig:syntax}.
Constraints make assertions about the structure of definitions and the inheritance graph.

Types $t$ are either type names~$\aeither$, variables $\tvar$, or method types $t \rightarrow t$.
Constraints, which may be conjoined by $\wedge$, take several forms.
A \emph{nominal subtyping} constraint $t_{1} \subtyp t_{2}$ holds if $t_{1}$ inherits from $t_{2}$ or implements $t_{2}$, possibly via inheritance.
The \emph{existential} constraint $\texists(\aeither)$ holds when the program contains a class or interface definition for $\aeither$.
Finally, we have two \emph{structural subtyping} constraints.
The constraint $t \subtyp \Struct{\meither : t_{M}}$ holds if $t$ defines or inherits a member $\meither$ of type $t_{M}$.
The constraint $\timpl(\aeither_{1}, \aeither_{2})$ holds if $\aeither_1$ defines or inherits members of the same name and compatible type for each member of $\aeither_{2}$.

Finally, constraints may also be \emph{suspended}, denoted $\Suspend{\Constraints}$.
We suspend constraints that are unsatisfied because they reference some undefined type or member.
For example, we might suspend $\texists(\aid)$ as $\Suspend{\texists(\aid)}$ if $\Prog$ currently contains no definition for $\aid$.
As discussed below in Section~\ref{subsec:synthesis}, during synthesis we unsuspend such constraints when we add to the program such that the constraints may become satisfied.
Our implementation also supports name constraints, which are discussed in Section~\ref{sec:implementation}.

\subsection{Type Inference for \Lm{}}
\label{subsec:typechecking}

Our synthesis algorithm uses type inference as a subroutine to generate and solve constraints among types and type variables.
\Cref{fig:constraint-generation} gives the type inference rules for \Lm{}.
The first several rules type expressions.
\textsc{TVar} types variables in the usual way. 
\textsc{TFld} types a field access $e.M$ by generating a constraint that the type $t$ of the object must contain a field~$M$ of type $\alpha$, where $\alpha$ is a fresh type variable.
\textsc{TCast} types a cast $(\aeither) e$ by generating two constraints: $\aeither \subtyp t$ requires that $A$ is a subtype of  $t$, and $\texists(\aeither)$ requires that $A$ exists in the program.
Notice that in a more typical type inference algorithm, these two conditions would be checked immediately, but in our system we need to generate constraints because the program may be partial.
Lastly, \textsc{TCall} types a method call $e_0.\meither(e)$ by constraining $t_0$, the type of the receiver, to contain a method that takes an argument of type $t$, the type of $e$, and returns some type $\alpha$.

\begin{figure}[t]
  \centering
  \footnotesize
  \begin{mathpar}
    \infer
        [TVar]
        { }
        { \CGEnv \vdash  x : \CGEnv(x) }

    \infer
        [TFld]
        { \CGEnv \vdash e : t  \\
          t \subtyp \Struct{ \meither : \alpha } \\
          \alpha \fresh
        }
        { \CGEnv \vdash e.\meither : \alpha }

    \infer
        [TCast]
        { \CGEnv \vdash e : t \\
          \aeither \subtyp t \\
          \texists(\aeither)
        }
        { \CGEnv \vdash (\aeither)\ e : \aeither}

    \infer
        [TCall]
        {
          \CGEnv \vdash e_0 : t_0 \\
          \CGEnv \vdash e : t \\
          t_0 \subtyp \Struct{ \meither : t \to \alpha } \\
          \alpha \fresh
        }
        { \CGEnv \vdash e_0.\meither(e) : \alpha }

    \infer
        [TMethD]
        {
          \ethis : \aeither_0, \superfg : \super(\Prog,\aeither_0),
          x : \aeither_2 \vdash e : t \\\\
          t \subtyp \aeither_1 \\ \texists(\aeither_1) \\ \texists(\aeither_2)
        }
        {
          \Prog; \aeither_0 \vdash \aeither_1\ M(\aeither_2\ x) \{ e \}
        }

  \infer
      [TFldD]
      { \texists(\aeither) }
      { \Prog \vdash \aeither\ \meither }

    \infer
        [TCls]
        {
          \Prog \vdash \ol{f} \\
          \Prog; \aeither_0 \vdash \ol{m} \\\\
          \texists(\aeither_1) \\
          \texists(\ol{\aeither}) \\
          \timpl(\aeither_0, \ol{\aeither}) \\
        }
        {
          \Prog \vdash \class\ \aeither_0 \extends \aeither_1 \implements
          \ol{\aeither}\ \{ \ol{f}\ \ol{m}\ \}
        }

  \infer
      [TIntf]
      {
        \texists(\ol{\aeither_{i0}}) \\
        \texists(\ol{\aeither_{i1}})
      }
      { \Prog \vdash \interface\ \aeither\ \{\ \ol{\aeither_{i0} \ \meither_i(\aeither_{i1}\ x_i)}\ \} }

  \end{mathpar}%

  \caption{Type Inference and Constraint Generation for \Lm{}.}%
  \label{fig:constraint-generation}
\end{figure}

\begin{figure*}[t]
 \FigSym{\ensuremath{\Prog \vdash \Constraints \becomesSymbol \Constraints'}}

\adjustbox{max totalheight=0.97\textheight}{
\footnotesize
\begin{tabular}{V|B}
  Nominals &
	\begin{constraints}[lcll]
      \con{\aeither \subtyp \aeither} \becomes{} \\
      \con{\aeither \subtyp \aeither'} \becomes[\Suspend{\aeither \subtyp
        \aeither'}] \cif{\aeither \notin \Prog} \\
      \con{\aeither \subtyp \aeither'} \fails{}
        \cif{\aeither \in \interfaces(\Prog) \text{ and } \aeither' \in \classes(\Prog)} \\
      \con{\aeither \subtyp \aeither'} \becomes{}
        \cif{\aeither \text{ inherits from or implements } \aeither'} \\
      \con{\aeither \subtyp \aeither'} \becomes[\Suspend{\aeither \subtyp
        \aeither'}] \cif{} \text{$\aeither$ inherits from or implements some $\aeither_i$ where $\aeither_{i}$ is a hole or $\aeither_{i} \notin \Prog$} \\
      \con{\aeither \subtyp \aeither'} \fails{} \cotherwise{}
    \end{constraints}
	\\

  Existentials &
	\begin{constraints}[lcll]
      \con{\texists(\aeither)} \becomes{} \cif{\aeither\in\Prog} \\
      \con{\texists(\aeither)} \becomes[\Suspend{\texists(\aeither)}] \cif{\aeither\notin\Prog}
    \end{constraints}
    \\

  Interfaces &
    \begin{constraints}[lcll]
      \con{\timpl(\aeither, \aeither')}
        \becomes[\timpl(\aeither, \aeither') \land \aeither \subtyp \Struct{ \meither : \aeither \to \aeither_r }]
        \cif{} \aeither'\in \interfaces(\Prog)
        \\ & & \cwhere{} (\aeither'.\meither : \aeither \to \aeither_r) \in \Prog \\
      \con{\timpl(\aeither, \aeither')} \fails{}
        \cif{} \aeither' \in \classes(\Prog) \\
      \con{\timpl(\aeither, \aeither')} \becomes[\Suspend{\timpl(\aeither,
        \aeither')}] \cotherwise{}
    \end{constraints}
	\\

  Variables &
    \begin{constraints}[lcll]
      \con{\aeither \subtyp \tvar \land \tvar \subtyp t}
        \becomes[\aeither \subtyp \tvar \land \tvar \subtyp t \land \aeither \subtyp t] \\
    \end{constraints}
	\\

  Methods &
    \begin{constraints}[rcll]
      \con{\aeither \subtyp \Struct{\meither : t \rightarrow t_r}}
        \becomes[t \subtyp t' \land t_r'
        \subtyp t_r]
        \cif{(\aeither.\meither : t' \rightarrow t_r')\in\Prog}
        \\
      \con{\aeither \subtyp \Struct{\meither : t \rightarrow t_r}}
        \becomes[t \subtyp t' \land t_r'
        \subtyp t_r]
        \cif{(\aeither'.\meither : t' \rightarrow t_r')\in\Prog}
        \\ & & \cwhere{\aeither \text{ inherits from }\aeither'}
        \\
      \con{\aeither \subtyp \Struct{\meither : t \rightarrow t_r}}
        \becomes[\Suspend{\aeither \subtyp \Struct{\meither : t \rightarrow t_r}}] \cotherwise{}
    \end{constraints}
\end{tabular}
}
\caption{Selected Constraint Resolution Rules.}%
\label{fig:constraint-resolution-rules}
\end{figure*}

The remaining rules type declarations.
\textsc{TMethD} types a method by ensuring that the method body $e$ meets its specification.
Notice that this rule generates $\texists$ constraints for the types mentioned in the method definition.
\textsc{TFldD} similarly generates an $\texists$ constraint for the field type.
\textsc{TCls} types a class definition by typing its fields and methods; generating $\texists$ constraints for the classes it mentioned (except $\aeither_0$, which is being defined); and generating $\timpl$ constraints to ensure $A_0$ is compatible with its superclass and interfaces.
Note that we abuse notation slightly here by writing $\texists(\ol{A})$ for the set of constraints $\bigwedge\texists(A_i)$ for $A_i \in \ol{A}$, and similarly for $\timpl(A_0, \ol{A})$.
Finally, \textsc{TIntf} types an interface by generating constraints that all classes mentioned in the interface exist.

\paragraph{Constraint Resolution}

After type inference generates constraints, constraint resolution iteratively rewrites the constraints until they are in a solved form.
In standard constraint resolution, constraints that are unsatisfiable always cause type inference to fail.
In contrast, in Mason, there are two possibilities:
\emph{Contradictory} constraints, which can never become satisfiable, force the synthesis algorithm to backtrack and try a different solution.
Otherwise, if the constraints could become satisfiable during synthesis, those constraints are suspended, and constraint resolution continues.

Figure~\ref{fig:constraint-resolution-rules} presents a selection of our constraint resolution rules, formalized as a relation $\Prog \vdash \Constraints \becomesSymbol \Constraints'$, meaning given program $\Prog$, the constraints $\Constraints$ are rewritten to $\Constraints'$.
We write $\becomesSymbol*$ for the reflexive, transitive closure of $\becomesSymbol$.
Contradictory constraints are those that are rewritten to $\bot$.
Note that we use the term \emph{implements} in the transitive sense; we say that $A$ implements $A'$ if $A \implements A'$ appears in source code or if one of $A$'s superclasses implements $A'$.

The nominal rewriting rules handle constraints of the form $\aeither \subtyp \aeither'$.
The first rule handles the case when $\aeither$ and $\aeither'$ are in $\Prog$ and the constraint is satisfiable, in which case it is discharged.
The next rule raises a contradiction if $\aeither$ is an interface and $\aeither'$ is a class.
The following two rules suspend the constraint when it is not currently satisfiable, but the inheritance or implementation chain includes a hole that is potentially on the path from $\aeither$ to $\aeither'$, since filling that hole could make this constraint satisfiable.

The existential rewriting rules either discharge a constraint $\texists(\aeither)$ if $\aeither\in\Prog$ or suspend the constraint, since synthesis could add $\aeither$ to the program.
The interface rewriting rules reason about $\timpl(\aeither, \aeither')$ constraints.
If $\aeither'$ is an interface in the program, then we generate new constraints that all the methods in $\aeither'$ are included in $\aeither$.
If $\aeither'$ is a class, then we have a contradiction.
Otherwise, we suspend the constraint until $\aeither'$ is added to the program.

The remaining rewriting rules are standard.
The variable rule implements transitive closure.

\subsection{Synthesis, Formally}
\label{subsec:synthesis}

Programs are synthesized from \emph{fragments}, which we define as a 2-tuple $\PS{}$ of a program $\Prog$ and constraints $\Constraints$.
Our goal is to derive, from $\PS{}$ and a set of fragments $F$ (a \emph{fragment library}), some fragment $\PSf[']$ where $\Constraints$ contains no suspended constraints.

The synthesis rules require a few key definitions.
A substitution $\sigma$ is a mapping of hole names to concrete type and member names.
If $\sigma = \{\ahole \mapsto \aid, \mhole \mapsto \methid\}$, then $\sigma(\PS{})$ denotes the program produced by replacing all instances of $\ahole$ with $\aid$ and $\mhole$ with $\methid$ in $\Prog$ and $\Constraints$.

We define a partial binary operator $\pcat{}$ on declarations.
$d_1 \pcat{} d_2$ is defined when $d_1$ and $d_2$ are of the same kind (class or interface), extend the same type, and contain no identically named members.
The result contains the union of the members of $d_1$ and $d_2$; if $d_1$ and $d_2$ are classes, the composition implements all interfaces implemented by either.
We lift $\pcat{}$ to programs by composing each pair of identically- named declarations; declarations with no counterpart are preserved in the result.
Finally, we lift $\pcat{}$ to fragments:
$\langle P, C \rangle \pcat{} \langle P', C' \rangle = \langle P \pcat{} P', C \land C' \rangle$.

\paragraph{Type- and Name-Guided Synthesis}
\Mason{} synthesizes programs by solving suspended constraints with \emph{type- and name-guided synthesis}, the rules of which we describe in \Cref{fig:synthesis-rules}.
Rules prove judgments of the form $\Fragments \vdash \PS{} \synthSymbol \PSf[']$ meaning that $\PSf[']$ can be synthesized from $\PS{}$ using fragment library $\Fragments$.

\begin{figure}[t]
  \centering
  \FigSym{\ensuremath{\Fragments \vdash \PS{} \synthSymbol \PSf[']}}
  \footnotesize
  \begin{mathpar}
    \inferrule[Fill-Type]
    {
      \aeither_1 \in \Prog \\
      \sigma(\aeither_0) = \sigma(\aeither_1) \\\\
      \langle \Prog', \Constraints_r \rangle = \sigma(\PS[\texists(\aeither_0)]) \\\\
      \Prog' \vdash \Constraints_r \becomesSymbol* \Constraints'
    }
    { \Fragments \vdash \PS[\Suspend{\texists(\aeither_0)}] \synthSymbol \PSf[']}

    \inferrule[Fill-Member]
    {  \aeither' \in \super^{*}(\Prog, \aeither) \cup \{ \aeither \} \\\\
      (\aeither'.\meither_1) \in \Prog \\
      \sigma(\meither_0) = \sigma(\meither_1) \\\\
      \langle \Prog', \Constraints_r \rangle = \PS[\aeither \subtyp \Struct{\meither_0 : t}]\\\\
      \Prog' \vdash \Constraints_r \becomesSymbol* \Constraints'
    }
    { \Fragments \vdash \PS[\Suspend{\aeither \subtyp \Struct{\meither_0 : t}}]
      \synthSymbol \PSf[']
    }

    \inferrule[Synth-Type]
    {
      \PSf{} \in \Fragments \\ \aeither_1 \in \Prog_f\\
      \sigma(\aeither_0) = \sigma(\aeither_1) \\\\
      \langle \Prog', \Constraints_r \rangle = \sigma(\PS[\texists(\aeither_0)]) \pcat{} \sigma(\PSf{}) \\\\
      \Prog' \vdash \Constraints_r \becomesSymbol* \Constraints'
    }
    {\Fragments \vdash \PS[\Suspend{\texists(\aeither_0)}] \synthSymbol \PSf[']}

    \inferrule[Synth-Member]
    {
      \aeither_0 \in \Prog \\ \PSf{} \in \Fragments \\
      (\aeither_1.\meither_1 : t_f) \in P_f \\\\
      \aeither_{s0} = \super(\Prog, \aeither_0) \\ \aeither_{s1} = \super(\Prog_f, \aeither_1) \\\\
      \sigma(\meither_0) = \sigma(\meither_1) \\
      \sigma(\aeither_0) = \sigma(\aeither_1) \\
      \sigma(\aeither_{s0}) = \sigma(\aeither_{s1}) \\\\
      \langle \Prog', \Constraints_r \rangle = \sigma(\PS[\aeither_0 \subtyp \Struct{\meither_0 : t}]) \pcat{} \sigma(\PSf{}) \\
      \Prog' \vdash \Constraints_r \becomesSymbol* \Constraints'
    }
    {
      \Fragments \vdash \PS[\Suspend{\aeither_0 \subtyp \Struct{\meither_0 : t}}]
      \synthSymbol \PSf[']
    }
  \end{mathpar}%
  \normalsize

  \caption{Synthesis Rules.}%
  \label{fig:synthesis-rules}
\end{figure}

Each synthesis rule matches on a suspended constraint in the fragment $\PS{}$; applies some transformation to $\PS{}$ to produce $\langle \Prog', \Constraints_{r} \rangle$; type checks the result by resolving $\Constraints_{r}$ in the context of $\Prog'$ to produce $\Constraints'$; and produces a new well-typed fragment $\PSf[']$ in which the constraint is discharged.
The \textsc{Fill-} rules do so by filling holes, while the \textsc{Synth-} rules do so by synthesizing new types and members.

Rule \textsc{Fill-Type} discharges a suspended existential constraint $\Suspend{\texists(\aeither_{0})}$ by unifying a hole with a concrete type name.
The first name, $\aeither_{0}$, is referenced in the constraint, while the other, $\aeither_{1}$, must occur as the name of a type declaration.
The rule defines $\sigma$ as the substitution unifying $\aeither_{0}$ and $\aeither_{1}$; recall that substitutions must map holes to concrete names.
It then applies $\sigma$ to $\PS{}$ to create a program in which a definition exists for $\sigma(\aeither_{0})$.

Rule \textsc{Fill-Member} discharges a suspended structural subtyping constraint $\Suspend{\aeither \subtyp \Struct{\meither_0 : t}}$ by unifiying $\meither_{0}$ with the name of one of $\aeither$'s members whose type is compatible with $t$.
Since $\aeither$ can inherit members from its supertypes, \textsc{Fill-Member} considers all member names from supertypes of $\aeither$ as candidates to unify with $\meither_{0}$.

Rule \textsc{Synth-Type} discharges a suspended existential constraint $\Suspend{\texists(\aeither_0)}$ by synthesizing fresh code.
The rule selects a fragment $\PSf{}$ from fragment library $\Fragments$ with a type declaration named $\aeither_{1}$ and uses $\sigma$ to unify it with $\aeither_{0}$ in both $\PS{}$ and $\PSf{}$.
It composes the resulting fragments to produce $\langle \Prog', \Constraints_{r} \rangle$, in which a definition now exists for $\sigma(\aeither_{0})$.

Rule \textsc{Synth-Member} discharges a suspended structural subtyping constraint $\Suspend{\aeither_0 \subtyp \Struct{\meither_0 : t}}$ by adding a new member to $\aeither_{0}$ whose type is compatible with $t$.
The rule selects a fragment $\PSf{}$ from $\Fragments$ and selects a method $\aeither_{1}.\meither_{1}$ from this fragment which to add to $\aeither_{0}$ using fragment composition.
To make merging $\aeither_{0}$ and $\aeither_{1}$ possible and to match the constraint, $\sigma$ unifies three pairs of names: method names $\meither_{0}$ and $\meither_{1}$; type names $\aeither_{0}$ and $\aeither_{1}$; the supertypes of both types $\aeither_{s0}$ and $\aeither_{s1}$.
Applying $\sigma$ to both fragments and composing them results in $\sigma(\aeither_{0})$ containing a definition for $\sigma(\meither_{0})$.

These rules are highly non-deterministic; given a set of suspended constraints, the rules may be applied in many different orders.
In \Cref{sec:implementation}, we will describe how the \Mason{} implementation systematically applies these rules to yield a search procedure.

\section{Implementation}
\label{sec:implementation}

\Mason{} is implemented in roughly 5,000 lines of Scala 3 code.
It takes as input an initial fragment containing unit tests and a library of fragments used to complete the initial fragment.
Both are written using a modified Java syntax that admits holes and name patterns.

The initial fragment cannot contain any holes, but is free to reference types and methods that do not yet exist.
We designate the class that declares the program's \texttt{main} method as its test harness. The \texttt{main} method may contain any number of unit tests, which are \texttt{static void} methods with the annotation \texttt{@Test}.
Unit tests may not rely on the result of any code executed outside their body, as trace-guided backtracking relies on this property during backtracking, and must raise an execution-halting exception to signal failure.

Although fragments are whole programs in the formalism, in the \Mason{}
implementation they are restricted to containing a single type declaration.
While it is useful to typecheck a fragment as a whole program, we have not found
a use for a multi-declaration fragment in representing any of our chosen design
patterns.

We also divide fragments between
type-fragments and member-fragments, i.e., between those which \Mason{} will
use to synthesize new types using \textsf{Synth-Type} and those which \Mason{}
will use to synthesize new methods using \textsf{Synth-Member}.
We denote this in the fragment source code using the type annotations
\texttt{@TypeFragment} and \texttt{@MemberFragment}.
This distinction prevents \Mason{} from wasting time by applying fragments in the wrong context.
For example, in \Cref{sec:overview}, the purpose of the fragment \texttt{?Update} shown in \Cref{fig:example-frag} is to synthesize setter methods on existing classes.
Using this fragment to synthesize a new class with a single member would uselessly expand the search space, so we designate \texttt{?Update} a member-fragment.

Finally, to prevent
unintended aliasing between holes introduced in different synthesis steps, the
\Mason{} implementation $\alpha$-renames holes in newly synthesized code.

\subsection{Synthesis, Practically}

In this subsection we describe the restrictions we impose to recover a search procedure from the highly non-deterministic synthesis rules described in \Cref{fig:synthesis-rules}.

\paragraph{Backtracking Search}
After parsing and typechecking both the initial fragment and the fragment
library, \Mason{} performs a depth-first backtracking search through the space of
synthesizable candidate programs. At each partial candidate program, it chooses a
suspended existential or structural subtyping constraint and attempts to
synthesize a new candidate program according to the rules in
\Cref{fig:synthesis-rules}. Recall we use two kinds of rules to solve subtyping constraints:  \textsc{Fill-} rules and \textsc{Synth-} rules.
We prefer to explore smaller candidates first, so \Mason{} will exhaustively explore
the space of candidates reachable by applying the \textsc{Fill-} rule before
doing the same for \textsc{Synth-}. If no solution is found here either,
\Mason{} backtracks to the previous candidate.

\paragraph{Constraint Ordering}
The order in which constraints are chosen affects the space of synthesizable programs.
If suspended constraints can be addressed in any order, \Mason{} may attempt to solve a constraint before it is possible to do so.
For example, suppose \Mason{} chooses a suspended constraint $\Suspend{\texttt{Foo} \subtyp \Struct { \texttt{f} : \texttt{Integer} }}$ given the following snippet:

\begin{lstlisting}
class Bar { Integer f; }
class Foo extends ?X { }
\end{lstlisting}

Two rules, \textsc{Fill-Member} and \textsc{Synth-Member}, match on this constraint.
Our first option here is to apply \textsc{Fill-Member} to concretize the name of
a field, but $\Foo$ has no eligible fields.
The second is to synthesize a new field with
\textsc{Synth-Member}, but what if our fragment library contains no
compatible member fragments? Instead, the solution here is to fill in
$\texttt{Foo}$'s superclass \texttt{?X} with \texttt{Bar}, making \texttt{f}
available through inheritance. However, \Mason{} will attempt the first two
strategies and backtrack, missing the solution.
As a result, it might only be possible to solve a constraint once a class has inherited particular members.

A variation of this dependency may also exist between two structural subtyping
constraints. Consider the following snippet and the constraints
$\Suspend{\texttt{Foo} \subtyp \Struct { \texttt{f} : \texttt{Integer} }}$ and
$\Suspend{\texttt{Bar} \subtyp \Struct { \texttt{f} : \texttt{Integer} }}$:

\begin{lstlisting}
class Bar { Integer ?f; }
class Foo extends Bar { }
\end{lstlisting}

Satisfying the constraint on \texttt{Bar} first concretizes the hole \texttt{?f}
and indirectly resolves the constraint on \texttt{Foo}. However, given that
no fragment exists to synthesize a new field, choosing the constraint on
\texttt{Foo} first leads to a failure, since nothing that can be done to
\texttt{Foo} will satisfy the constraint.

As a third example, consider a situation in which a constraint containing a
hole can only be satisfied once a constraint containing a concrete
name is satisfied. Consider the constraints
$\Suspend{\texttt{Foo} \subtyp \Struct { \texttt{f} : \texttt{Integer} }}$ and
$\Suspend{\texttt{Foo} \subtyp \Struct { \texttt{?g} : \texttt{Integer} }}$ and
following snippet:

\begin{lstlisting}
class Bar {
  Integer foo() { return this.?g }
  Integer ?h;
}
\end{lstlisting}

If \Mason{} chooses to satisfy
$\Suspend{\texttt{Foo} \subtyp \Struct { \texttt{f} : \texttt{Integer} }}$
first, we can fill \texttt{?h} with \texttt{f} in this step and then satisfy
$\Suspend{\texttt{Foo} \subtyp \Struct { \texttt{?g} : \texttt{Integer} }}$ by
filling \texttt{?g} with \texttt{f}. However, addressing the constraints in the
opposite order fails, since our rules do not allow us to unify \texttt{?g} and
\texttt{?h}. A similar situation may arise if
$\Suspend{\texttt{Foo} \subtyp \Struct { \texttt{f} : \texttt{Integer} }}$ is
satisfied by synthesizing a new field.

For these three kinds of dependencies we impose three rules. First, existential
constraints are always satisfied before subtyping constraints. Second,
constraints on supertypes always precede those on subclasses. Third,
for pairs of constraints not already comparable via the first two rules, a
constraint containing no holes always precedes a constraint containing some.

As a result, if a concrete name for a type or member exists in the current constraint set, \Mason{} will synthesize its definition before filling any holes for which it is a solution.
Regardless of ordering, if a concrete-named definition for a type or member exists in a fragment, it will be available for use via the \textsf{Synth-} rules.
Still, if the correct substitute for a hole exists only in a constraint within a library fragment, \Mason{} may attempt to fill the hole before first pulling in the fragment and be unable to solve it.
Thus, our constraint ordering guaratees that \Mason{} will find a solution as long as the name of every synthesized type and member in the solution is given either as a concrete reference in the initial fragment or as the name of a definition in a library fragment.

\subsection{Extensions}
\label{sec:extensions}

We now describe two extensions to \Mason{} aimed at mitigating combinatorial
explosion, particularly in cases where \Mason{} must choose between a number of
type-correct options which differ only in their runtime behavior.

\paragraph{Name Patterns}
Synthesis in \Mason{} revolves around filling holes with concrete names, thus
we can prune a great deal of the search space by imposing a naming convention to
which candidate substitutions must adhere. As described in
\Cref{sec:overview}, Mason lets the programmer  encode such a convention using
name patterns.

Name patterns may appear in source code in place of type or member names and
enclosed by \texttt{?\{\_\}}. They use a restricted regular expression syntax
with the addition of holes prefixed by \texttt{?}. For the purposes of this
paper we retain only the choice \texttt{|}, grouping \texttt{()}, and implicit
concatenation operators.
Name patterns are compiled to regular expressions by replacing every alphanumeric
sequence prefixed by \texttt{?} with \texttt{.*} and all uppercase letters with
their lowercase counterpart.

For each type name pattern \texttt{?\{}$\patc$\texttt{\}} or member name pattern
\texttt{?\{}$\patm$\texttt{\}} that appears in the source, \Mason{} generates a
corresponding \textit{name constraint} $\meither \matchm \patm$ or
$\aeither \matcha \patc$, where $\meither$, $\aeither$, is the entire
\texttt{?\{\_\}}-enclosed sequence interpreted as a type or member name. A name
constraint is satisfied when its left-hand side is a concrete name matched by
the pattern on its right-hand side.

The holes contained within a name pattern are renamed with the rest of the
fragment during synthesis. For example, the substitution
$\{ \texttt{?field} \mapsto \texttt{name} \}$
applied to pattern \texttt{(set)(?field)} yields \texttt{(set)(name)}.
Name patterns can encode syntactic relationships between
various holes, e.g., the relationship between the name of a setter and the
name of the field it mutates, as was shown in \Cref{sec:overview}.

\paragraph{Trace-Guided Backtracking}
\Mason{} implements a non-local backtracking heuristic, \emph{trace-guided backtracking}, which we briefly described in \Cref{sec:overview}.
To enable this, \Mason{} tracks an \emph{interference set} for every synthesis step taken and a \emph{dependency set} for every unit test run.

The \emph{interference set} of a synthesis step tracks the set of methods whose implementation may depend on its outcome.
It begins empty for each step and is populated with the methods that synthesis rules \emph{interfere} with as they are attempted.

If \textsc{Fill-Member} is used to rename a hole $\mhole$, it interferes with the following methods:

\begin{itemize}
  \item any member in which $\mhole$ appeared prior to renaming
  \item if this step gave a concrete name to a member named $\mhole$ in type $t$,
        any member with a hole name that \emph{ever exists} (exists now or is synthesized later) in $t$.
\end{itemize}

The second item is necessary because giving a particular name to one method in a
type declaration prevents that name from ever being given to any other method currently
in the declaration or ever added to the declaration in the future.

If \textsc{Synth-Member} is used to synthesize a method $\methid$ in type $t$, it interferes with the following:

\begin{itemize}
  \item any member in which $\mhole$ appeared prior to renaming
  \item any new member synthesized during this step
  \item any member with a hole name that ever exists in $t$, its subclasses, or
        its supertypes
\end{itemize}

If a synthesis step applies \textsc{Fill-Class} or \textsc{Synth-Class}, the
interference set is every method in the program.

The \emph{dependency set} of a unit test tracks the set of methods upon whose correctness the outcome of the unit test might depend.
When any unit test fails, the set of methods called during its execution are added to the test's dependency set.
We further expand the dependency set to include the names of any methods prior to being renamed by \textsc{Fill-Member}.

When \Mason{} encounters a unit test failure, the dependency set for that test is
updated and \Mason{} backtracks non-locally until reaching a synthesis step
whose interference set intersects with the dependency set.
Search resumes without non-local backtracking, although if \Mason{} visits every outgoing edge
without finding a executable program, non-local backtracking resumes using the same
dependency set. During this time, \Mason{} prioritizes any choices made while synthesizing the last executable program.

It is possible that in attempting to correct for a failure of some unit test $T_{2}$,
\Mason{} makes a change that causes a previously passing test $T_{1}$ to fail.
In this case, we backtrack and search for a new solution with the
intent of passing both $T_{1}$ and $T_{2}$.
\Mason{} must revise its choices at steps that affect both tests, so the dependency set of $T_{1}$ becomes the union of the dependency sets of $T_{1}$ and $T_{2}$. The execution trace, and trace-guided backtracking, proceed as usual.

\section{Evaluation}
\label{sec:evaluation}

We evaluate \Mason{} on a suite of benchmarks that involve synthesizing programs using common design patterns.
The questions we aim to answer in our evaluation are:

\begin{enumerate}
  \item[RQ1] How well does \Mason{} achieve its stated goal of automatically
        synthesizing implementations of design patterns?
        How do particular aspects of a synthesis problem cause \Mason{} to
        perform well or poorly?
  \item[RQ2] To what extent and in which scenarios are the two extensions to \Mason{}
        effective in improving its performance?
\end{enumerate}

\begin{figure*} [ht]
  \includegraphics*[width=14cm]{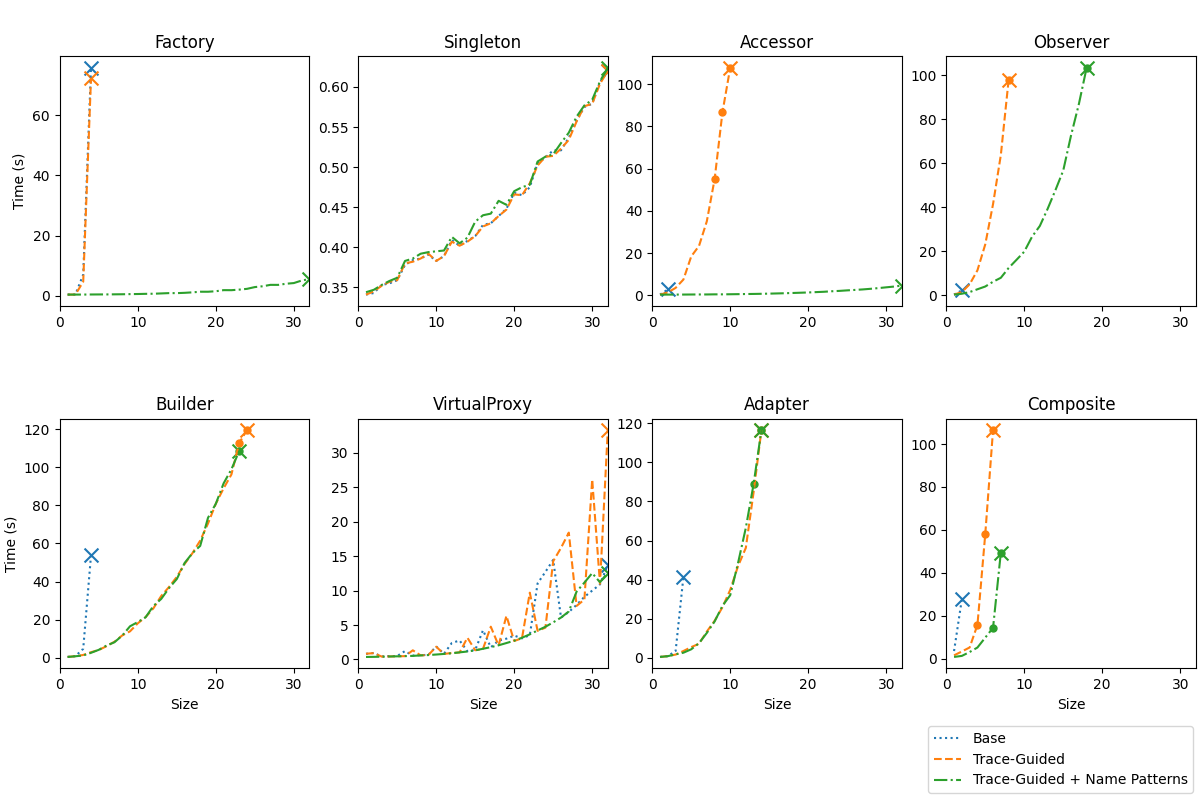}
  \caption{
    Benchmark size versus median wall time over 21 trials for each of our 8 benchmarks.
    Note that y-axis scales vary between subplots.
    We stop plotting points once the median trial exceeds the timeout.
    The last data point for each series of tests is marked with an $\times$.
    Runs where at least one trial timed out are marked with a $\bullet$.
  }
  \label{fig:benchmark-results}
\end{figure*}

\subsection{Benchmarks}
We developed our own synthetic benchmark suite designed to reveal \Mason{}'s strengths and weaknesses.
In each benchmark, \Mason{} synthesizes a program by adding a Gang of Four \cite{gamma1994} design pattern, which we have implemented as a fragment library.
The initial fragment for each benchmark includes a set of unit tests that instantiate classes and check that the design pattern was implemented correctly.
For example, the unit tests for the \emph{Accessor} benchmark check that setters mutate the correct fields and that these mutations are correctly observed by getters.
The size of each benchmark can be adjusted by a scaling factor $n$, so that we can evaluate Mason's performance as benchmark size grows.
To give a sense of scale, \Mason{} synthesizes 298 additional lines of code (as measured by \texttt{cloc}~\cite{cloc}) to complete the \emph{VirtualProxy} benchmark at size 32.
We now describe each benchmark.

\begin{itemize}
  \item
        \textbf{Factory.} The initial fragment consists of $n$ \emph{product} classes that implement a common interface.
        For each, \Mason{} must synthesize a \emph{concrete factory} that instantiates objects of that class.
        \Mason{} must also synthesize an \emph{abstract factory} interface that the concrete factories implement.
        The fragment library contains type fragments for synthesizing concrete factory classes and the abstract factory interface.
  \item
        \textbf{Singleton.} To each of $n$ classes in the initial fragment, \Mason{} must add a field \texttt{\_instance} and a method that calls the class's constructor to initialize \texttt{\_instance} if it is uninitialized, returning its value otherwise.
        The fragment library contains a member fragment that includes both the field and method.
  \item
        \textbf{Accessor.} The initial fragment contains a class with $n$ fields, all of the same type.
        For each, \Mason{} must synthesize setters and getters.
        The fragment library contains a member fragment for synthesizing setters and another for getters.
  \item
        \textbf{Observer.} This benchmark generalizes the example from \Cref{sec:overview}.
        Along with $n$ setters on a \emph{subject} class, \Mason{} must synthesize a fixed number of helper methods for managing observers.
        The fragment library contains a member fragment containing the helper methods and the fragment shown in \Cref{fig:example-frag}, for setters.
  \item
        \textbf{Builder.} The initial fragment contains a class with $n$ constructor parameters.
        \Mason{} synthesizes a \emph{concrete builder} for this class, which has a setter for each parameter and a method for instantiating the class.
        The fragment library contains a class fragment for synthesizing the builder, a member fragment for setters, and member fragments which instantiate classes.
  \item
        \textbf{Virtual Proxy.}  \Mason{} must synthesize a \emph{virtual proxy}, which exposes the same interface as a given class with $n$ methods but redirects calls to an instance of the class that it instantiates lazily.
        The fragment library contains type fragments for synthesizing the proxy class and method fragments for proxy methods.
  \item
        \textbf{Adapter.} \Mason{} must synthesize an \emph{adapter} that wraps an object of a given class with $n$ methods, while implementing an incompatible interface by redirecting method calls to the wrapped object.
        The fragment library contains type fragments for synthesizing the adapter class and method fragments for synthesizing its methods.
  \item
        \textbf{Composite.} Given $2n$ classes that define a tree structure, \Mason{} must synthesize a shared interface and implementations of it for each class, allowing a client to perform the same operation across leaves and non-leaves.
        The fragment library contains a type fragment and member fragment for synthesizing and populating the shared interface, a member fragment with ``non-leaf'' methods and another with ``leaf'' methods.
\end{itemize}

\subsection{Results}
\Cref{fig:benchmark-results} shows the results of running \Mason{} on our
 benchmark suite. The data was collected on an 2023 Macbook Pro
with an M2 Pro processor and 16GB of RAM. The benchmarks themselves were run on
a JVM with a maximum of 1GB of stack space and 8GB of heap.

For each benchmark, we ran and plotted three series of tests. The first, our baseline, referred to in the legend as \emph{Base}, were performed without trace-guided backtracking or name constraints.
The second, labelled \emph{Trace-Guided}, runs with only trace-guided backtracking switched on.
The third, labelled \emph{Trace-Guided + Name Patterns}, runs with both extensions on.

In each series, we instantiated the benchmark at progressively higher
scaling factors, beginning at one and increasing by increments of one up to and including 32.
At each scaling factor we ran 21 trials with a timeout of 120 seconds and recorded the total wall clock time spent in synthesis and unit test execution before all unit tests passed.
We then plotted the median of these times. If a majority of the 21 trials resulted
in a timeout, we did not run tests for any larger scaling factors and did not plot
a value.

\paragraph{Baseline}
Baseline \Mason{} performs well when a well-typed candidate program tends to be a correct one.
This is the case for the \emph{Singleton} benchmark, in which the special constructor for each class is synthesized using a single member fragment, both of whose holes are filled immediately by \textsc{Synth-Member}.
Likewise, for \emph{VirtualProxy}, there are only ever two possible well-typed candidates, in one of which the virtual proxy erroneously wraps an object of its own class.
As a result, synthesis times do not exceed 0.7 seconds for Singleton, and 15 seconds for VirtualProxy.

For the remaining benchmarks, increasing scale vastly increases the number of well-typed synthesizable programs, and performance suffers as a result.
For example, in the \emph{Accessor} benchmark, baseline \Mason{} times out for a majority of trials at a scale factor of three.
This makes sense when we examine the benchmark; for \emph{Accessor} at scale factor three, \Mason{} chooses, for each of three setter and three getter methods, the correct field to access, for a total of $3^{3} \times 3^{3}$ or 729 possible programs.

\paragraph{With Trace-Guided Backtracking}
With the addition of trace-guided backtracking, \Mason{} is able to synthesize correct solutions within the timeout at larger scale factors for \emph{Accessor}, \emph{Observer}, \emph{Builder}, \emph{Adapter}, and \emph{Composite}.
Like the \emph{Observer} example from \Cref{sec:overview}, the source of combinatorial explosion in these benchmarks is a series of choices made when synthesizing disjoint methods.

This is most pronounced in \emph{Builder}, where \Mason{} with trace-guided backtracking reaches size 24 and encounters no timeouts through size 22.
Baseline \Mason{} times out in a majority of trials after size four.
Though outperforming the baseline, \Mason{} does not fare as well in \emph{Composite}: here, \Mason{} is more prone to make changes that cause previously passing tests to fail, which expands dependency sets, making trace-guided backtracking less effective.
Note too that the initial \emph{Composite} program is larger than the others, containing $2n$ incomplete classes.

\paragraph{With Name Patterns}
When \Mason{} is also permitted to use the \emph{name constraints} generated from
\emph{name patterns} embedded in fragments, we see a noticable improvement in
performance in four benchmarks: \emph{Factory}, \emph{Accessor}, \emph{Observer},
and \emph{Composite}. Of these, \emph{Factory} and \emph{Accessor} see the most benefit,
with median synthesis times for \emph{Factory} never exceeding six seconds and
median times for \emph{Accessor} never exceeding five seconds.

This is somewhat expected; at the synthesis steps which cause combinatorial explosion in both benchmarks, name constraints force a single option.
In \emph{Accessor} this is the choice of field to manipulate in each setter and getter (similar to the example from \Cref{sec:overview}), and in \emph{Factory} this is the choice of class to instantiate in each concrete factory.
For \emph{Observer} and \emph{Composite}, name constraints fail to fully constrain every important choice, but still eliminate a significant amount of unnecessary search:
in \emph{Observer} at size eight, name constraints bring median synthesis times from 98 seconds to 13 seconds;
in \emph{Composite} at size six, they bring median synthesis times from 107 seconds to 14 seconds.

\section{Future Work}

There are several interesting directions for future work.

\paragraph{Name Patterns}
Name patterns caused the most drastic performance improvements in \Cref{fig:benchmark-results}, specifically with the \emph{Factory} and \emph{Accessor} benchmarks.
However, the language of name patterns is limited, and they can only be used to express constraints between type and member names.
Future work could extend name patterns to refer to names found in other parts of the AST, e.g., the names of method parameters.

\paragraph{Synthesizing Low-Level Details}
With \Mason{}'s higher level of abstraction comes a loss of expressiveness.
For instance, \Mason{} is unable to synthesize abritrary expressions or sequences of statements.
Future work might consider integrating \Mason{} with existing program synthesis techniques to make this possible.
We might permit holes to appear in expression position, with \Mason{}
producing \emph{sketches}~\cite{solarlezama09} to be solved using a tool like
\textsc{JSketch}~\cite{jeon2015}.

\paragraph{More Expressive Specifications}
Our evaluation demonstrates that \Mason{} performs particularly well when static information
tightly constrains the search space. However, Java types are a fairly coarse
specification.
LiquidJava~\cite{liquidjava} extends the Java type system with statically checked
predicates.
Future work could extend \Mason{}'s system of type
constraints to include these richer types and therefore prune more candidates during search.

\section{Related Work}

\paragraph{Synthesis of Object-Oriented Programs}
\textsc{Pasket}~\cite{jeon2016} also synthesizes Java program using design patterns.
However, while \Mason{}'s output is intended for use by programmers, \textsc{Pasket} produces framework models for symbolic execution.
Using logs produced by example framework clents, \textsc{Pasket} instantiates design patterns to produce a \emph{sketch}~\cite{solarlezama09} of the solution, which it completes using a tool called \textsc{JSketch}~\cite{jeon2015}.
Like fragments, sketches are partial programs with holes, but holes range over expressions rather than names.
To solve sketches, \textsc{JSketch} reasons about dynamic semantics but not static semantics, thus \textsc{JSketch} and \textsc{Pasket} may produce ill-typed programs, while type-guided \Mason{} will not.
A consequence of reasoning about dynamic semantics is that \textsc{JSketch} and \textsc{Pasket} cannot synthesize code that uses library methods without a semantic model; meanwhile, \Mason{} only needs type signatures, which are available by default.

Various techniques~\cite{guria2021, guria2023, mariano19, feng17} exist to synthesize individual functions in object-oriented languages like Ruby and Java.
These techniques synthesize expressions or sequences of statements that \Mason{} cannot but lack the ability to create new classes, interfaces, and members.
Of these, \textsc{RbSyn}, which synthesizes Ruby functions from API calls, shares similarities with \Mason{}.
\textsc{RbSyn} also uses unit tests and test failures to guide search, though it does so by requiring candidate programs to perform additional side ffects, which \textsc{RbSyn} reasons about directly.
However, \textsc{RbSyn} requires that library methods include effect annotations as a result.

\paragraph{Type-Directed Synthesis}
\textsc{Synquid}~\cite{polikarpova2016}, \textsc{Myth}~\cite{osera2015}, and \textsc{Scythe}~\cite{osera2019} represent the traditional approach to type-directed synthesis, deriving synthesis rules from proof systems for typechecking and type inference.
In contrast, though type inference is a key part of synthesis in \Mason{}, \Mason{}'s synthesis rules do not follow its structure.
In addition, all three synthesize individual functions or expressions in a functional language.
Like \Mason{}, \textsc{Synquid} guides synthesis with subtyping constraints, but uses more expressive \emph{refinement types} which are annotated with logical predicates.
Like \Mason{}, \textsc{Myth} also uses test cases to guide synthesis, but these are less expressive input-output examples.
However, \textsc{Myth} is able to refine its examples alongside synthesis, using them to eliminate candidates during search.
\textsc{Scythe} carries a set of typing constraints through synthesis for a similar reason to \Mason{}: in the context of type inference, filling in parts of a program may have non-local implications.
However, \textsc{Scythe} reasons only about equality constraints.

\paragraph{Design Pattern Automation}
A related topic in the software engineering literature is automatic refactoring of code to use design patterns.
In contrast to \Mason{}, these techniques are applied to complete programs, and are thus intended for use later in the development process.
\textsc{Remodel}~\cite{jensen10} leverages a genetic algorithm to apply \emph{minitransformations}~\cite{cinneide00}, small code changes that can compose design patterns, and determines fitness using software engineering metrics.
\textsc{JungGL}~\cite{jungl} and $\mathcal{R}^{2}$~\cite{kim15} are tools for writing \emph{refactoring scripts}, which apply design patterns using a specialized API.
Like \Mason{}'s fragments, scripts allow user configuration, but authors of refactoring scripts must devise an algorithm, while \Mason{} obviates this need.
Producing well-typed programs is also a concern in this domain; Tip et al. \cite{tip03} use typing constraints to ensure the well-typeness of refactorings.
Like us, they define a system of typing constraints for a Java subset and use them to constrain possible program transformations.
However their treatment of constraints is specific to each refactoring: some constraints must hold following the transformation, while the rest need not.

\section{Conclusion}

In this paper, we presented \Mason{}, a tool that, guided by types and names, synthesizes object-oriented
programs from program pieces. We gave a formal model
of our approach, introducing a constraint-based type inference algorithm for a
core object-oriented language and a set of type- and name-guided synthesis rules.
We also introduced two extensions to \Mason{} with the goal of improving performance.

We used \Mason{} to apply design patterns to a suite of benchmarks.
We found baseline \Mason{} to perform well when very few well-typed programs can be synthesized.
When this is not the case, but the source of combinatorial explosion is distributed among distinct methods and unit tests, trace-guided backtracking can improve performance over baseline \Mason{} significantly.
Name patterns further improve performance when a naming convention can be imposed on types and members.
We believe that \Mason{} takes an important step in synthesizing multi-class object-oriented programs using design patterns.

\section*{Data Availability Statement}

We have made the source code for \Mason{}, the benchmark suite used in the paper, scripts and instructions to reproduce our results from \Cref{sec:evaluation}, and a Docker image available as an artifact~\cite{artifact}.

\bibliography{paper}

\end{document}